# Deep Learning based Modeling of Wireless Communication Channel with Fading

Youngmin Lee, Xiaomin Ma, Andrew S.I.D. Lang, Enrique F. Valderrama-Araya, Andrew L. Chapuis
College of Science and Engineering, Oral Roberts University, Tulsa, OK 74171, USA
Emails: ttmn_ym@oru.edu, xma@oru.edu, alang@oru.edu, evalderrama@oru.edu, and chapuis_andrew@oru.edu

*Abstract*— In the realm of wireless communication, stochastic modeling of channels is instrumental for the assessment and design of operational systems. Deep learning neural networks (DLNN), including generative adversarial networks (GANs), are being used to approximate wireless Orthogonal frequency-division multiplexing (OFDM) channels with fading and noise, using real measurement data. These models primarily focus on channel output (y) distribution given input x: p(y|x), limiting their application scope. DLNN channel models have been tested predominantly on simple simulated channels. In this paper, we build both GANs and feedforward neural networks (FNN) to approximate a more general channel model, which is represented by a conditional probability density function (PDF) of receiving signal or power of node receiving power $P_{rx}$: $f_{P_{rx}|d}()$, where is communication distance. The stochastic models are trained and tested for the impact of fading channels on transmissions of OFDM QAM modulated signal and transmissions of general signal regardless of modulations. New metrics are proposed for evaluation of modeling accuracy and comparisons of the GAN-based model with the FNN-based model. Extensive experiments on *Nakagami* fading channel show accuracy and the effectiveness of the approaches.

## I. INTRODUCTION

Modern wireless communication systems play a pivotal role in connecting individuals, supporting emerging technologies, and driving the digital transformation of societies worldwide. In wireless communication systems, a signal transmitted from one node to other node(s) could be severely decayed or attenuated by fading and shadowing, path loss caused by long communication distance, big building blocks, and interferences from other transmitting nodes. It is critical to consider the effect of the channel when a communication system is designed [1], and its quality of service (QoS) is analyzed for the optimization of the parameters or network configuration [2]-[4]. Since the practical channels are very complicated and vary with time and location, it is hard to obtain the channel transfer function in advance. Most previous analyses and designs assume theoretical expressions to characterize the effect of the channel on the communication systems. Recently, several pure data-driven approaches to channel modeling using deep learning neural networks have been proposed [5]-[9]. Soltani et al. [6] proposed a deep learning algorithm for channel estimation in communication systems by considering the time-frequency response of a fast-fading communication channel as a 2D image. Xiao et al. [7] designed a ChannelGAN algorithm on a small set of 3rd generation partnerships project (3GPP) link-level multiple-input multiple-output (MIMO) channel. O'Shea et al. [8] approximated the channel output (y) distribution given channel input x: p(y|x) using variational Generative Adversarial Networks (GANs). This stochastic channel model was improved to a conditional GAN (cGAN) by incorporating the received pilot information as additional information [9] and applied to the optimal design of end-to-end learning-based communication systems [10]. Since the DLNN models only consider the fading effect without accounting for the impact of path loss effect on the transmitting signal in the channel, the application scope of this model is somewhat limited. So far, these models have been applied to classification of modulated signals in relatively simple channels such as the additive white Gaussian noise (AWGN) channel, the Rayleigh fading channel, or similar [9]-[10]. Furthermore, lower convergency speed and poor robustness of GANs hinders these approaches from being used for real-time applications.

The objective of this paper is to leverage two deep learning models, GANs and feedforward neural networks (FNN), to approximate stochastic characteristics of a more general channel model. Compared with the existing methods for channel modeling, the main contributions of this paper are: 1) The proposed deep learning model is more general, which can produce channel models for the purpose of both classification and regression and account for effect of fading and communication distance. 2) Comparisons of different DLNN models and structures for modeling of *Nakagami* fading channel with path loss are conducted. The DLNN models proposed in this paper are more precise and robust. 3) New metrics are proposed for objective evaluation of modeling accuracy and comparisons of DLNN channel models. The paper is organized as follows. Section II gives a brief overview of wireless communication systems, channels, and channel modeling. Section III describes channel modeling based on GANs and FNNs, respectively. Section IV illustrates the setup of the experiments to verify the accuracy and effectiveness of the DLNN modeling scheme. Then, the numerical results of the proposed scheme and discussions are given. Conclusions are presented in Section V.

## II. SYSTEM DESCRIPTION

### A. Wireless Communication Systems and Channels

Many wireless communication systems have been developed to enable seamless communication across diverse platforms with high quality of services (QoS). The communication systems include, but are not limited to, upgraded Wi-Fi networks for high-speed internet access, ad hoc networks for mission-critical applications, and 5G/6G Cellular networks for personal communications. These systems leverage innovative technologies such as advanced coding methods in the physical layer (e.g., low-density parity checking (LDPC)), high speed modulation and coding schemes (MCS) (e.g., quadrature amplitude modulation (QAM) 1024), Multi-user Multiple-input Multiple-output (MU-MIMO) and orthogonal frequency division multiplexing access (OFDMA), etc.

Wireless communication channels are characterized by various impairments, including fading, shadowing, path loss, noise, and interferences from other nodes' transmissions. Fading refers to the rapid variations in signal strength due to factors like multipath propagation. Shadowing occurs due to obstacles in the signal path, leading to signal attenuation in specific areas. Path loss represents the gradual reduction in signal strength over distance, influenced by factors such as free-space loss and absorption by atmospheric gases. Noise and interferences take place when nodes receive multiple signals from different transmitters or other sources in the same channel. Understanding and mitigating these challenges are crucial for analysis and design of robust wireless systems.

*B. Objectives and Formulation of Channel Modeling*

To characterize a general wireless channel, we configure a deep learning-based model that is trained using collected data from measurement at both senders and receivers. As shown in Fig. 1, inputs of the channel to be modeled are either signal transmitting signal strength $x$ or sending signal power $P_t$, and outputs of the channel are either receiving signal strength y or receiving signal power $P_r$. The distance between the sender and the receiver is denoted as $d$.

The objective of the modeling is to train the deep learning neural networks to approximate channel statistical impacts on the transmitting signals, which are characterized by the cumulative distribution function (CDF) and the probability density function (PDF) of power or strength of signal at a receiver with distance $d$ away from a source node:

$$F_{P_r|d}(y) = \Pr(P_r \leq y|d), \quad (1)$$

$$f_{P_r|d}(y) = \frac{dF_{P_r|d}(y)}{dy}, \quad (2)$$

and

$$\overline{P_r(d)} = PL(P_t, d) + N_G, \quad (3)$$

where $\overline{P_r(d)}$ is average receiving signal strength or power as a path loss function $PL(P_t, d)$ of transmitting power and communication distance d. $N_G$ is added noise.

Two approaches are proposed to generate data pairs to train the neural networks. Approach 1 is a simple and natural extension of the method in [9] where the labeling model is built and tested for the purpose of QAM signal classification. While Approach 2 is a more general way to model wireless channels, in which regression model is configured and tested for real values of signal strength and power.

**Approach 1**: Collect or generate data pairs of $(x_i, y_i)=(P_t, P_r)$ given a distance $d$ and $2^M$ input signals with M-QAM modulation are assumed.

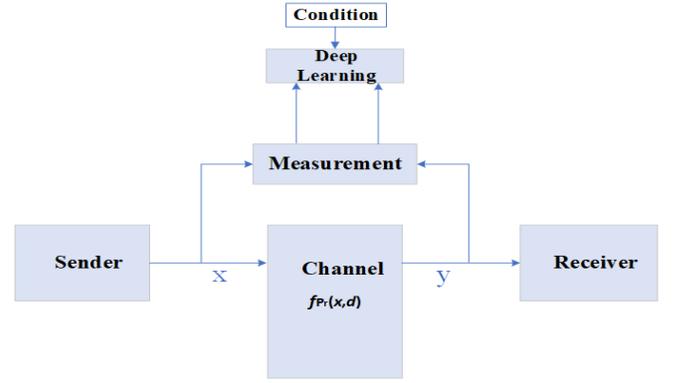

Fig. 1 Structure of deep learning-based channel modeling

a) Define M-QAM input signals $P_t^i = A_i cos(2\pi f + \Phi_i)$, i=1, 2,…, $2^M$, each signal represents one $M$ bit message;
b) Given a fixed distance d, generate a random number $A_i$ and $\Phi_i$ (i=1,2,…, $2^M$), apply the input signal $P_t^i$ to the channel to be investigated;
c) Measure $P_r$ and estimate CDF: $F_{P_r}(x)$;
d) Repeat a)~c) until the number of data pairs is sufficient;
e) i = i+1, i ≤ $2^M$ repeat a)~d).

**Approach 2**: Collect or generate data pairs of $(x_i, y_i)=(d, P_r)$ without considering type of modulation.
a) Decide ranges of modulated signal parameters $P_t$ ;
b) Given two nodes with distance value $d_i$, apply the input signal $P_t$ to the channel to be investigated;
c) Measure $P_r$ and distance $d_i$;
d) Estimate CDF: $F_{P_r|d}(y)$ and PDF: $f_{P_r|d}(y)$;
e) Repeat a)~d) until the number of data pairs is sufficient;
f) Find other node pairs, repeat a)~e).

III. MODELING APPROACHES

*A. Conditional GAN for Channel Evaluation*

The first deep learning model we adopt for the channel modeling is a GAN that can learn a model producing samples close to some target distribution. The conditional GAN model in [9] is adopted and extended to the modeling of more general channel characteristics and regression distribution of signal strength and signal power. The objective function of cGAN learning is expressed as [9]

$$\min_G \max_D = V(D, G) =$$
$$E_{x \sim p_{data}(x)}[logD(x|c)] + E_{z \sim p_z(z)}\left[\log\left(1 - D(G(z|c))\right)\right]. \quad (4)$$

From Eq. (4), we can see that cGAN model focuses on maintaining the equilibrium between Generator and Discriminator loss values while learning against each other [11]. In other words, cGAN's primary object is not a regression task. Also, the Generator from the cGAN used Gaussian Error Linear Units (GELU), which has been verified as having better performance than Rectified Linear Units (ReLU) in general cases [12].

$$GELU(x) = xP(X \leq x) = x\Phi(x) = x \cdot \frac{1}{2}[1 + \text{erf}\left(\frac{x}{\sqrt{2}}\right)] \quad (5)$$

The pseudo code of the cGAN for channel modeling is shown below.

**Algorithm 1** Build Model - cGAN

**Input:** $x_i$ genuine data, $z_i$ noise, $c_i$ condition
**Output:** generated data $G(z_i|c_i)$
1: **BUILD** Generator $G$ by following
2:    input1 = Input($z_i$)
3:    input2 = Input($c_i$)
4:    x = Concatenate(input1, input2)
5:    **FOR** num_layer
6:      x = Dense(num_unit, Activation=Gelu)(x)
7:      x = BatchNormalization()
8:    **END FOR**
9:    x = Dense(2)(x)
10: **BUILD** Discriminator $D$ by following
11:    input1 = Input($G(z_i|c_i)$ and $x_i$)
12:    input2 = Input($c_i$ and $c_i$)
13:    x = Concatenate(input1, input2)
14:    **FOR** num_layer
15:      x = Dense(num_unit, activation=Gelu)(x)
16:    **END FOR**
17:    x = Dense(1, Activation=Sigmoid)(x)
18: **SET** $D$ untrainable
19: **BUILD** cGAN by combining $G$ and $D$ with Eq. (4)
20: **COMPILE** cGAN and $D$ with Optimizer=Adam, Loss=BinaryCrossentropy, Metrics=Accuracy

---

**Algorithm 2** Build Model - FNN

**Input:** genuine data $x_i$, condition $c_i$
**Output:** generated data $FNN(c_i)$
1: **BUILD** FNN by following
2:    x = Input($c_i$)
3:    **FOR** num_layer
4:      x = Dense(num_unit, activation=Gelu)(x)
5:      x = BatchNormalization()
6:    **END FOR**
7:    x = Dense(2)(x)
8: **COMPILE** FNN with Optimizer=Adam, Loss=MSE or RMSE, Metrics=[MSE and RMSE]

## B. Multi-layer Feedforward Deep Learning for Channel Modeling

The first deep learning model we adopt for the channel modeling is a feedforward neural network (FNN). The structure and training of the FNN is described in Algorithm 2 pseudo code.

$$MSE = \frac{1}{n}\sum_{i=1}^{n}(real_i - predicted_i)^2. \quad (6)$$

$$RMSE = \sqrt{\frac{1}{n}\sum_{i=1}^{n}(real_i - predicted_i)^2}. \quad (7)$$

Since FNN models in this research accepted *mean squared error* (MSE, Eq. (6)) and *root mean squared error* (RMSE, Eq. (7)), they can conduct a regression task efficiently. Especially since RMSE (and MSE) is recommended over than *mean absolute error* (MAE) when the error forms a normal distribution like ours; $N_G$ (Gaussian noise) [13].

## IV. NUMERICAL RESULTS AND DISCUSSIONS

### A. Setup of Experiments for Testing and Comparisons

In this research, we choose to use a *Nakagami* channel with exponential path loss specifically for the experiment because it fits the practical wireless fading channel well and Rayleigh channels are special cases of *Nakagami* channels. The channel modeling and the evaluation were done for two scenarios: Approach 1 when $x$ is signal voltage magnitude. Approach 2 when $x$ is power in watts. In addition, Approach 1 is basically the reproduction of H. Ye *et al.*'s works [9]-[10] extended into our *Nakagami* channel instead of one of the Rayleigh channels.

Eq. (8) below is to calculate PDF of strength $P_{rx}$ conditioned on communication distance $d$

$$f_{P_{rx}|d}() = \frac{1}{\Gamma(m)}\begin{cases}\left(\frac{m}{\overline{P_r(d)}}\right)^m x^{m-1} exp\left(-\frac{mx}{\overline{P_r(d)}}\right); \text{ if Approach 1}\\ 2\left(\frac{m}{\overline{P_r(d)}}\right)^m x^{2m-1} exp\left(-\frac{mx^2}{\overline{P_r(d)}}\right); \text{ if Approach 2}\end{cases}, \quad (8)$$

where $\Gamma(m)$ is the Gamma function, and $m$ is the fading parameter, and

$$\overline{P_r(d)} = PL(P_t, d) = P_t\eta\left(\frac{d_0}{d}\right)^\alpha. \quad (9)$$

Then, the Cumulative Distribution Function (CDF) of $P_{rx|d}$ can be expressed as

$$F_{P_{rx}|d}() = \begin{cases}\frac{1}{\Gamma(m)}\int_0^{\frac{mx}{\overline{P_r(d)}}} t^{m-1}e^{-t}dt = \frac{\gamma\left(m,\frac{mx}{\overline{P_r(d)}}\right)}{\Gamma(m)}; \text{ if Approach 1}\\ \frac{1}{\Gamma(m)}\int_0^{\frac{mx^2}{\overline{P_r(d)}}} t^{m-1}e^{-t}dt = \frac{\gamma\left(m,\frac{mx^2}{\overline{P_r(d)}}\right)}{\Gamma(m)}; \text{ if Approach 2}\end{cases}. \quad (10)$$

where $\gamma(s,y) = \int_0^y t^{s-1}e^{-t}dt$ is the lower incomplete gamma function. Ultimately, the random value of $P_r$ can be calculated with the inverse function of CDF or the lower incomplete gamma function.

$$P_r(d) = \begin{cases}\left(\frac{\overline{P_r(d)}}{m}\gamma^{-1}(m,\Gamma(m)r)\right)^{\frac{1}{2}}; \text{ if Approach 1}\\ \frac{\overline{P_r(d)}}{m}\gamma^{-1}(m,\Gamma(m)r); \text{ if Approach 2}\end{cases}. \quad (11)$$

The normal Gaussian distribution function is adopted to generate the channel noise $N_G$ with the mean of the distribution $N_{G_{mean}}$ and the variance $N_{G_{var}}$.

In Approach 1, the authors set 4-QAM modulations {-3, -1, 1, 3} to define 16 input signals for $P_t$. The rest of the parameters' values are: $m = 1, \eta = 1, d_0 = 100, d = 200$, and $\alpha = 2$. The total data for Approach 1 experiment is 1.08 million values with an 8:2 ratio of train and validation, and another 0.1 million for the test.

In Approach 2, while $P_t$ is fixed to $0.28183815W$, $d$ has 30 different categories ($d = \{10, 20, 30, ..., 300\}$), and $m$ has two categories ($m = 2$ if $d \le 140$ else $m = 1$). The rest of the variables are: $\eta = 7.29 \times 10^{-14}, d_0 = 100, \alpha = 2$.

Approach 2 trains cGANs and FNNs (MSE and RMSE loss) with three different data sizes of 0.54, 1.08, and 2.16 million values without noise. Then applied the Gaussian noise to the channel while $N_{G_{mean}} = 1.256 \times 10^{-15}W$, and $N_{G_{var}} = 1 \times 10^{-15}$ and $1 \times 10^{-16}$ with the fixed 1.08 million data size.

In the experiment, $\eta, N_{G_{mean}}$, and $N_{G_{var}}$ are scaled with $10^{14}$ for convenience. We also applied log scaling (Eq. (12)) for $P_r(d)$ before feeding them to the DLNNs for the

**Algorithm 3** Data Preparation

**If Approach 1**
  **Input:** $m, \eta, d_0, d, \alpha, r_i$ $(0 \leq r_i \leq 1), N_{G_i}, P_{t_{array}}$
**If Approach 2**
  **Input:** $P_t, \eta, d_0, \alpha, r_i$ $(0 \leq r_i \leq 1), N_{G_i}, m_{array}, d_{array}$
**End IF**
**Output:** genuine data $x_i$, condition $c_i$
1: **FOR** num_data
2:   **GENERATE** genuine data $x_i$
3:     $x_i$ = Eq. (11) + $N_{G_i}$
4:   **GENERATE** condition $c_i$
5:   **IF Approach 1**
      # can substitute Normalize with Embed
6:     $c_i$ = Concatenate(Normalize($P_{t_i}$), $r_i$)
7:   **IF Approach 2**
      # if using Embed, this should be inside of the DLNN
8:     $c_i$ = Concatenate(Embed($d_i$), $r_i$)
9:   **END IF**
10: **END FOR**

following reasons: 1) The distribution of $P_r(d)$ is positively Skewed; 2) and its range varies dramatically with $d$. Then, the values were scaled back after the DLNN models.

$$scaled\ x = \begin{cases} \log_{10}(x) & ;if\ w/\ noise \\ \log_{10}(x+2) & ;if\ w/o\ noise \end{cases}. \quad (12)$$

The authors trained the cGAN for 50000 epochs and saved the model every 100 epochs and trained the FNN for 500 epochs and saved after each epoch. We have not considered overfitting on FNN because the training, validation, and test datasets are generated by the same equation Eq. (8).

Therefore, we propose and establish new metrics to evaluate and compare the performance of the modeling in an objective way.

*Scaled Percent Error* (*ScaledPE*) to identify the best model performance. For each model with epochs, calculate the mean and variance of each ideal data (Eq. (13)-(14)) and generated data (Eq.(15)-(16)) for every category.

$$IdealMean(P_r(d)) = \begin{cases} \frac{\Gamma(m+\frac{1}{2})}{\Gamma(m)} \left(\frac{\bar{P}_r(d)}{m}\right)^{\frac{1}{2}} ; if\ Approach\ 1 \\ \bar{P}_r(d) \quad ; if\ Approach\ 2 \end{cases} \quad (13)$$

$$IdealVar(P_r(d)) = \begin{cases} (\bar{P}_r(d)) \left(1 - \frac{1}{m}\left(\frac{\Gamma(m+\frac{1}{2})}{\Gamma(m)}\right)^2\right) ; if\ Approach\ 1 \\ \frac{\bar{P}_r(d)^2}{m} \quad ; if\ Approach\ 2 \end{cases} \quad (14)$$

$GeneratedMean(P_r(d))$, and $GeneratedVar(P_r(d))$ where $n$ is the number of $P_r(d)$ for each category ($P_{t_{array}}$ or $d_{array}$).

$$\bar{x} = \frac{\Sigma x}{n} \quad (15)$$

$$s^2 = \frac{\Sigma(X-\bar{x})^2}{n-1} \quad (16)$$

*PEMean*, and *PEVar* the percent error for the mean and variance between the ideal and generated data for each category.

$$PE = \left|\frac{Ideal - Generated}{Ideal}\right| \times 100. \quad (17)$$

*ScaledPE* indicates the average Scaled Percent Error if it does not specify its categorical value. which considers the variance more significant. (Other ratio values would be possible instead of 0.3 and 0.7.)

$$ScaledPE = \frac{(PEMeanAvgh * 0.3 + PEVarAvg * 0.7)}{2}. \quad (18)$$

Finally, select the model that has the lowest *ScaledPE*. In this experiment, the authors used a total of 0.3 million data values to calculate the *ScaledPE*.

*Overlapped Area* (*OA*), which exhibits how the PDF of receiving signal from the DLNN model synchronizes with the ideal PDF (with noise). The process to determine the overlapped area is: 1) Generate *genuine data* from Eq. (11) (apply the noise if needed) and *generated data* from the DLNN by valid categories ($P_{t_{array}}$ or $d_{array}$). 2) Calculate the Gaussian Kernel Density Estimation for both *genuine* and *generated data* for each category.

$$f_\sigma(x) = \frac{1}{n\sigma} \sum_{i=1}^{n} K\left(\frac{x-x_i}{\sigma}\right). \quad (19)$$

We chose $K$ for the Gaussian kernel and $\sigma = 0.3$ in this experiment. 3) Determine the smaller PDF values using Eq. (20), where $\{x_i\}_{i=1}^{i=k}$ is a partition or $[s_{min}, s_{max}]$ with $x_1 = s_{min}$ and $x_k = s_{max}$. We defined $k$ as the number of $P_r(d)$ for each category.

$$f_{lower}(x_i) = Min\left(f_{\sigma\sim genuine}(x_i), f_{\sigma\sim generated}(x_i)\right). \quad (20)$$

where $i = 1, 2, ..., k$

Approximate the *local Overlapped Area* using the trapezoidal rule.

$$OA_{local} = \sum_{i=1}^{k-1} \frac{1}{2}(f_{lower}(x_i) + f_{lower}(x_{i+1}))\Delta x, \quad (21)$$

where *local* refers to each category while $\Delta x = \frac{s_{max}-s_{min}}{k}$.

Then, calculate the average *OA* of the entire categories to generalize the DLNN model's overall performance with Eq. (16) while $n$ is the length of the categories. This paper's term *OA* indicates the average of $OA_{local}$ if not specified. In this experiment, we generated 0.1 million $P_r(d)$ to calculate *OA*. Also, higher *OA* implies better PDF accordance in genuine and generated data.

In short, the authors selected the best cGAN and FNN with *ScaledPE,* which can generate the most accordant data compared to the ideal data. Then, used *OA* to compare the performance of each model with an additional perspective.

### B. Numerical Results and Discussions

Our experiment using Approach 1 shows that the FNN (with MSE) model outperformed the cGAN model. According to Fig. 2, while the best FNN model on Approach 1 has *ScaledPE* of 2.07 and 1.62 for real and imaginary pair, cGAN has 8.11 and 6.64 which are about 4 times bigger than that in FNN. This result implies that $P_r(d)$ generated by FNN behaves more like to its ideal mean and variance than cGAN. FNN also yields better *OA* value pairs (0.96 and 0.96) than cGAN (0.86, 0.81). In other words, the probability of the imaginary $P_r(d)$ occurrence by its power scale between the ideal (genuine, through Eq. (11)) and the generated through FNN has 96% of correspondence. However, cGAN only reaches to 81%. The constellation of $P_r(d)$ by cGAN also expresses the unsatisfaction of its performance (Fig. 3). In addition, the constellations by FNN of all QAM-modulations are closer to the ideal coordination and the entire $P_r(d)$, which corresponds to their $P_t$, converge at one point (Fig.4).

In Approach 2, like in Approach 1, we found that FNN

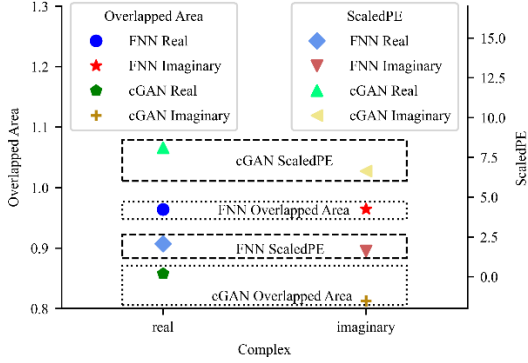

Fig. 2: (Approach 1) *Overlapped Area* between the genuine and generated (left) and *ScaledPE* (right) of each real and imaginary pair of 4-QAM modulations. The generated data came from FNN (mse) and cGAN trained with 1.08 million data values.

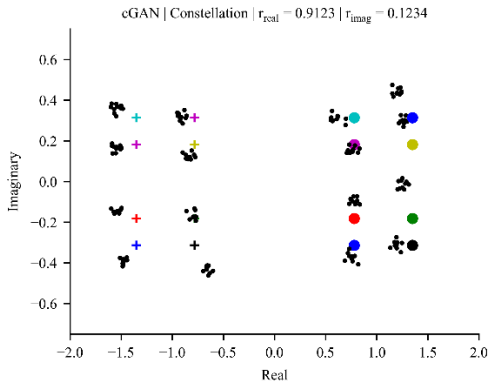

Fig. 3: (Approach 1) Constellation of 4-QAM modulations generated by cGAN while $r_{real} = 0.9123$ and $r_{imaginary} = 0.1234$.

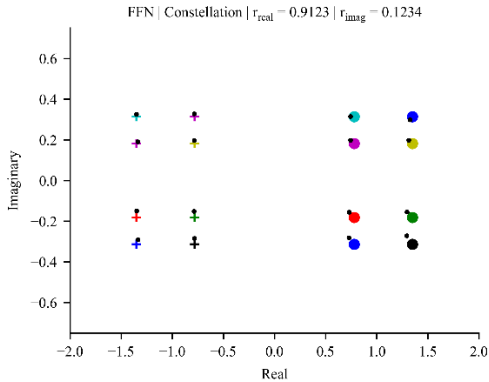

Fig. 4: (Approach 1) Constellation of 4-QAM modulations generated by FNN while $r_{real} = 0.9123$ and $r_{imaginary} = 0.1234$.

models over-perform cGAN models in general. When noise is not applied to the channel, then both *ScaledPE* and Overlapped Area from FNNs surpass the other as shown in Fig. 7. Also, both *ScaledPE* and *OA* values do not change by the data size of 0.54, 1.08, and 2.16 millions, respectively.

However, FNN models do not always beat cGAN if the noise is applied. When $N_{G_{var}} = 1 \times 10^{-15}$, although *ScaledPE* on the FNNs is better than that for cGANs, the *OA* values over the FNN models and the cGAN models are not distinguishable. According to Fig. 9, for example, when $N_{G_{var}} = 1 \times 10^{-15}$, the *ScaledPE* values of FNN with *MSE* = 3.26 and with *RMSE* = 3.67 are considerably lower than that of cGAN's (=9.55). However, *OA* of all three DLNN models looks very similar to each other (FNN$_{\text{mse}}$: 0.84, FNN$_{\text{rmse}}$: 0.85, and cGAN: 0.83).

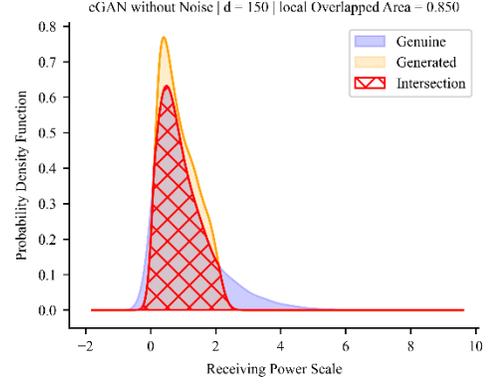

Fig. 5: (Approach 2) *OA* of $P_r(d)$'s PDF between the genuine data calculated by Eq. (11) and the generated data from cGAN when the distance $d = 150m$ and $N_{G_{var}} = 1 \times 10^{-16}$.

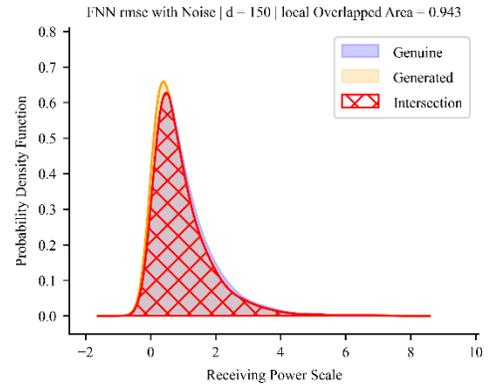

Fig. 6: (Approach 2) *OA* of $P_r(d)$'s PDF between the genuine data and the generated data from FFN (rmse) when $d = 150m$ and $N_{G_{var}} = 1 \times 10^{-16}$.

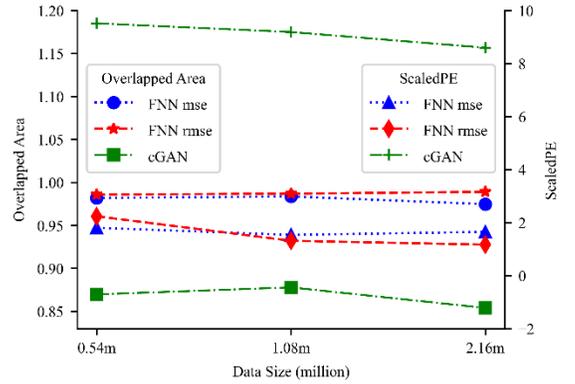

Fig. 7: (Approach 2) *OA* (left) and *ScaledPE* (right) among FNN with *MSE* & *RMSE* and cGAN trained with different data sizes.

More specifically, FNN models start from a higher *OA* with a smaller distance $d$, then rapidly decrease its value to the greater $d$. The cGAN model begins with a lower value but declines with a lower negative slope (Fig. 9). In addition, if noise is not applied, FNN keeps all *OA* values by $d$ close to the average *OA* without dramatic change.

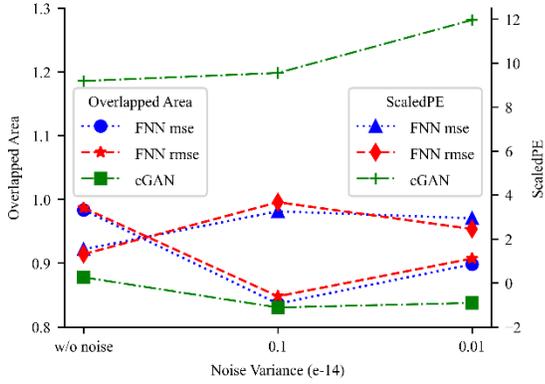
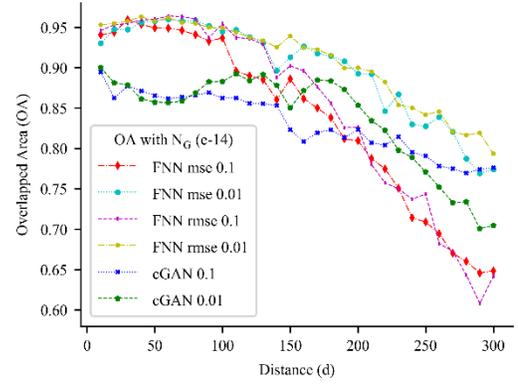

Fig. 8: (Approach 2) *OA* and *ScaledPE* among FNNs and cGAN trained with 1.08 million data w/o noise, $N_{G_{var}} = 1 \times 10^{-15}$, and $N_{G_{var}} = 1 \times 10^{-16}$.

Fig. 9: local *OA* (not average *OA*) by each distance $d$ from FNNs and cGAN with different $N_{G_{var}}$.

On the other hand, FNNs with $N_{G_{var}} = 1 \times 10^{-16}$ outran cGAN overall again with both *ScaledPE* and *OA*. Figs. 5-6 display the performance of FNN (RMSE) and cGAN in the perspective of *OA*. More specifically, FNN with RMSE has about 94% of PDF~$P_r(d)$ accordance between the genuine (which is Eq. (11) + $N_G$) and the generated data. However, cGAN has about 85% of area correspondence. Also, the FNN model has lower absolute average slope from the first and the last *OA* than cGAN. Another point to remark from the experiment is that the FNN with RMSE loss function performs slightly better than *MSE* in general.

In other words, FNN noticeably outperformed cGAN through the experiment in general. However, when the range of noise distribution gets bigger, the PDF generated by cGAN shapes more like its genuine PDF than FNN when the distance increases. Nonetheless, if there is no noise or its range of distribution is small, FNN models explicitly surpass cGAN. This result corresponds to the characteristics of the loss functions that the FFNs and the cGAN adopted. When the features and targets have a strong correlation (i.e., small or no $N_{G_{var}}$), FNNs can handle this effectively as a conventional regression task. While cGAN with the loss function Eq. (7) can become a substitute if the covariates and outcomes pair have a weak correlation (i.e., big $N_{G_{var}}$ and $P_r(d)$ gets faded by a long distance $d$).

## V. CONCLUSIONS

In this paper, we apply two deep learning neural networks to the modeling of wireless channels. Also, we introduce new metrics and evaluate and compare modeling approaches: *ScaledPE* and *OA*. Through the experiments on *Nakagami* fading channel with exponential path loss, we observe that FNN outperformed cGAN in *ScaledPE* (mean and variance of $P_r(d)$) with all the scenarios. However, OA reveals that PDF of $P_r(d)$ generated by cGAN can coincide more precisely with that of the original data when the noise variance is huge and if the noise becomes relatively significant at receiver end. Therefore, if only considering the mean and variance (or our *ScaledPE*) of $P_r(d)$, FNN will be one of the best options for both Approach 1 and Approach 2. However, if considering *OA*, then cGAN also can be a better choice when $N_{G_{var}}$ is big or the impact of $N_G$ becomes relatively more considerable at receiver end. Our future research will focus on modeling more general channels.

REFERENCES

[1] Q. Mao, F. Hu, and Q. Hao, "Deep learning for intelligent wireless networks: A comprehensive survey," *IEEE Communications Surveys and Tutorials*, 20(4), 2595–2621, April 2018.
[2] J. Zhao, Z. Wu, Y. Wang and X. Ma, "Adaptive optimization of QoS constraint transmission capacity of VANET," *Vehicular Communications*, vol. 17, pp. 1-9, June 2019.
[3] S. Ding and X. Ma, "Model-based deep learning optimization of IEEE 802.11 VANETs for safety applications," *International Wireless Communications and Mobile Computing (IWCMC)*, Dubrovnik, Croatia, 2022.
[4] X. Ma and K. Trivedi, "SINR-based analysis of IEEE 802.11p/bd broadcast VANETs for safety services," *IEEE Transactions on Network and Service Management*, 18(3): 2672-2686. March 2021.
[5] T. J. O'Shea, and J. Hoydis, "An introduction to deep learning for the physical layer," *IEEE Transactions on Cognitive Communications and Networking*, vol. 3, no. 4, pp. 563 – 575, Oct. 2017.
[6] M. Soltani, V. Pourahmadi, A. Mirzaei, and H. Sheikhzadeh, "Deep learning-based channel estimation," *IEEE Communications Letters*, 23(4): pp. 652-655, April 2019.
[7] H. Xiao, *et al.*, "ChannelGAN: Deep learning-based channel modeling and generating," *IEEE Communications Letters*, 11(3): pp. 650-654, March 2022.
[8] T. J. O'Shea, T. Roy, and N. West, "Approximating the void: learning stochastic channel models from observation with variational generative adversarial networks," *IEEE International Conference on Computing, Networking and Communications* (*ICNC*), Honolulu, USA, April 11, 2019.
[9] H. Ye, G. Li, B. F. Juang, and K. Sivanesan, "Channel agnostic end-to-end Learning based communication systems with conditional GAN," *IEEE Globecom Workshops*, Abu Dhabi, UAE, Dec. 9-13, 2018.
[10] H. Ye, G. Li, and B. Juang, "Deep learning based end-to-end wireless communication systems without pilots," *IEEE Transactions on Cognitive Communications and Networking*, vol. 7, no. 3, pp. 702 – 714, Feb. 2021.
[11] M. Mirza, and S. Osindero, "Conditional Generative Adversarial Net," arXiv:1411.1784[cs.LG], Nov. 2014.
[12] D. Hendrycks, and K. Gimpel, "Gaussian Error Linear Units (GELUs)," arXiv:1606.08415[cs.LG], Jun. 2016.
[13] T. Chai, and R. Deaxler, "Root mean square error (RMSE) or mean absolute error (MAE)?," *Geoscientific Model Development,* vol. 7, no. 3, pp. 1247-1250, Jun. 2014.